\documentclass[18pt]{article}
\usepackage{bbm}
\usepackage[final]{graphicx}
\usepackage{amsmath}
\usepackage{amsfonts,amsbsy}
\usepackage{amssymb}
\usepackage[breaklinks,dvipdfm]{hyperref}
\usepackage{subfigure}
\usepackage{graphicx}

\begin{document}

\begin{flushright}
YITP-SB-11/47 \\
\end{flushright}

\centerline{\Large Partial fractioning reduction of perturbative amplitudes}

\vbox{\vskip .5 in}

\begin{center}
{Stanislav Srednyak \\ C. N. Yang Institute for Theoretical Physics \\
Stony Brook University, Stony Brook, New York 11794-3840, USA}
\end{center}

%\begin{abstract}
%A novel method is presented for the reduction of loop integrals in a field theory with large numbers of external lines.  As the number of loops $L$ is kept fixed, an arbitrary diagram reduces to a finite basis of intergrals. The number of denominators in the basic integrals grows as a multiple of $L-1$, with a coefficient that depends on the theory and dimension, independent of the number of external lines.
%\end{abstract}

\begin{abstract}
A new  method is presented for the simplification of loop integrals in one particle irreducible diagrams with large numbers of external lines, based on the partial fractioning of products of propagators. Whenever a loop diagram in $d$ dimensions has $d+1$ or more lines that carry the same linear combination of loop momenta, its integral can be reexpressed as a linear combination of integrals with no more than $d+1$ denominators for each such set of lines, of which $d$ are linear in the loop momenta and only one quadratic.   In multiloop diagrams, the total number of linear denominators can be reduced further.    In integrals with  numerator momenta there may also be up to $d+1$ linear factors in the numerator. 
\end{abstract}

\section{Introduction.}
Tree and loop amplitudes and cross sections with multiple external lines in gauge and other field theories are the subject of  wide attention ~\cite{Britto:2010xq}, ~\cite{Dixon:2011xs},  ~\cite{Binoth:2010ra}. Much of this  work is based on analyticity properties of the integrals ~\cite{hwa},~\cite{Argeri:2007up}, and on the identification of bases for integrals with fixed number of denominators ~\cite{Laporta:2001dd}.

 If the number of loops is kept fixed, one might expect that the diagram function of a field theory, in some dimension $d$, would get progressively more complicated as the number of external lines is increased. For one loop, however, it has long been known that the number of  denominators that are quadratic in the loop momentum can be reduced to no more than five for each loop ~\cite{Melrose:1965kb} in four dimensions.   This paper presents an alternative method for reducing a diagram with any number of loops and many external lines to a  linear combination of integrals of limited complexity. These integrals correspond to vacuum bubble diagrams with the same number of loops as the original diagram and with up to $d +1$ denominators per line, $d$  of which are linear in the loop momenta. The number of such diagrams increases linearly in $L-1$, independent of the number of external lines. Such integrals have $d+2$ poles per loop, to be compared to $2(d+2)$  poles in the integral of a standard diagram. The method of Ref. ~\cite{vanNeerven:1983vr},\cite{Bern:1992em} can then be applied to derive a further reduction to $d$ denominators, at least in the one loop case.  The reduction of the number of lines for single-particle irreducible diagrams is analogous to the recursive structure of gauge theory tree amplitudes, ~\cite{Britto:2005fq} and the underlying graphical identities are in fact related  ~\cite{Vaman:2005dt,Draggiotis:2005wq}.

The reduction is based on two fairly simple identities, which are presented in Section 2.   The general formula for the class of what will be called ``sufficiently dressed" diagrams is also presented there. For such diagrams the total number of denominators that depend on loop momenta $q_i$ is $k(d+1)L$, with $k$ a number that is defined by the theory and does not grow with $d$ or $L$. For instance, for $\phi^3$ theory $k$ equals  $3$.   Of these denominators, $k\times d\times L$ are linear, the others quadratic in $q$.  In Section 3 it is shown how to extend the method to theories with spinors and vectors. The general result is that there is a basis for the integrals such that for each set of lines carrying a specific linear combination of loop momenta there is only one quadratic and up to $d$ linear factors in the denominator and up to $d+1$ linear factors in the numerator.  It is possible to further reduce the number of linear denominators. The method for doing this is illustrated by a two-loop example in scalar theory in Section 4. 

\section{Identities.}

We consider scalar integrals in $d$ dimensions. To each diagram, we associate a vacuum bubble diagram, found by removing all external lines. In this section the method is presented in the  case, when the diagram is ``sufficiently dressed''. A diagram is sufficiently dressed if there are at least $d$  external lines emanating from each line in the corresponding bubble diagram.  The reduction is performed for  arbitrary complex masses and complex external momenta. 

In the diagram, consider the set of denominators that corresponds to a given linear combination of loop momenta, $q$. This set of lines can be identified with a single line in the corresponding bubble diagram. The starting point of the reduction is the following identity, applied to any such set of lines,
\begin{align}
\prod_{e=1}^{N_q} \frac{1}{(q+p_e)^2+m_e^2} =  \sum_{e=1,...,N_q} \frac{1}{(q+p_e)^2+m_e^2}  \label{eq:first}\\ \nonumber
& \ & \hspace{-50mm}
\times \prod_{j=1,...,N_q,j\neq e} \frac{1}{2q\cdot(p_j-p_e)+p_j^2+m_j^2-p_e^2-m_e^2}\ ,
\end{align}
where $N_q$ is the number of external lines attached to $q$, and $p_e$ denotes the sum of external momenta carried on line $e$. This identity is valid as long as none of the denominators vanishes. The  $q,p_e$ are $d$-component momenta, and $m_e$ are the (possibly complex) masses of the lines. Masses can be zero. The integral is regulated by off-shellness. This equation is valid for all $N_q>1$. 

Equation ~\eqref{eq:first} is a consequence of the following identity which will be used repeatedly below,
\begin{align}
\prod_{e=1}^N \frac{1}{s_ex+a_e}=\sum_{e=1}^N\frac{1}{s_ex+a_e} \prod_{f=1,...,N,f \neq e} \frac{1}{s_fx+a_f} |_{x=X^e}, \quad s_eX^e+a_e=0\ .  \label{eq:Identity}
\end{align}
In this equation, $s_e,x,a_e$ are arbitrary complex numbers, considered for a generic point in the space $\{s_e,a_e,x\}$. The identity ~\eqref{eq:Identity} is most easily proved by considering the integral
\begin{align}
Int=\frac{1}{2\pi i}\int_C \frac{d\xi}{\xi} \prod_{e=1}^N \frac{1}{s_e(x+\xi)+a_e}\ , \label{eq:Int}
\end{align}
where the contour of integration $C$ encircles zero in a counter-clockwise sense.  The integral over such contour is equal to the left hand side of Eq\ .~\eqref{eq:Identity}. When a circle at infinity is added to such a contour, the integration can be closed on all the other poles of the integrand, giving in this way the sum of residues of the integrand other than from zero. This is equal to the right hand side of Eq\ .~\eqref{eq:Identity}.

In particular, setting $s_e=s$ for all $e$  and taking the limit $s\rightarrow 0$, the identity reduces to
\begin{align}
\prod _{e=1}^N\frac{1}{a_e}=\sum_{e=1}^N \frac{1}{a_e}\prod_{f=1,...,N,\neq e}\frac{1}{a_f-a_e}\ .
\end{align}
Taking $a_e=(q_e+p_e)^2+m_e^2$ gives the identity Eq\ .~\eqref{eq:first}.

We next assume that $N_q>d$. By further partial fractioning the product of linear factors in ~\eqref{eq:first} can be further reduced, by using an identity whose proof will be given momentarily. For arbitrary $N_q>d+1$ the identity is
\begin{align}
\prod_{j\neq e}^{N_q} \frac{1}{2q\cdot(p_j-p_e)+p_j^2+m_j^2-p_e^2-m_e^2} & = & \label{eq:second} \\ \nonumber
& \ & \hspace{-50mm}
\sum_{j_1\neq ... \neq j_d \neq e} ( \prod_{s=1...d}\frac{1}{2q\cdot(p_{j_s}-p_e)+p_{j_s}^2+m_{j_s}^2-p_e^2-m_e^2}   ) T^e_{j_1,...,j_d}(p_{e'})\ , 
\end{align}
where the factors $T^e_{j_1,...,j_d}(p_{e'})$ are functions of external momenta only,
\begin{align}
T^e_{j_1,...,j_d}(p_{e'})=\prod_{k\neq j_1,...,j_d,e} \frac{1}{2Q_{j_1,...,j_d}\cdot(p_k-p_e)+p_k^2+m_k^2-p_e^2-m_e^2}\label{eq:T}\ .
\end{align}
In this expression, the $d$-dimensional  vectors $Q_{j_1,...,j_d}$ are the solutions of the following system of $d$ linear equations, corresponding to the ``missing'' denominators in each of the terms on the right-hand side of Eq.\ (2),
\begin{align}
2Q\cdot(p_{j_1}-p_i)+p_{j_1}^2+m_{j_1}^2-p_e^2-m_e^2=0 \label{eq:system}\\ \nonumber
....  \\ \nonumber
2Q\cdot(p_{j_d}-p_i)+p_{j_d}^2+m_{j_d}^2-p_e^2-m_e^2=0\ . \\ \nonumber
\end{align}
Thus the argument of $T^e_{j_1,..., j_s}(p_{e'})$ represents masses and all momenta $p_{e'}$ that do not appear in the $q$-dependent  denominators of that term. This sytem is to be solved at generic values of momenta and masses, i.e. symbolically. It is worth noting that solutions to Eq.\ \eqref{eq:system} are analogous to the complex momenta that set internal lines of the S-matrix on shell in Ref.\ \cite{Britto:2005fq} and discussed diargammatically in Refs. ~\cite{Vaman:2005dt,Draggiotis:2005wq}.

The proof of ~\eqref{eq:second} again relies on the identity ~\eqref{eq:Identity}. In order to simplify the notation, we rewrite the identity to be proven in the form
\begin{align}
F_N(\{l\},\{a\},q)=\prod_{e=1}^N \frac{1}{l_e\cdot q+a_e}=\label{eq:tobeproven} \\ \nonumber
& \   & \hspace{-50mm}
\sum_{e_d\neq e_{d-1}\neq...\neq e_1} \prod_{s=1,...,d}\frac{1}{l_{e_s}\cdot q+a_{e_s}} T_{e_d,...,e_1}(\{l\},\{a\})\ , 
\end{align}
where $q\in C^d, l_e \in C^d, a_e \in C$ and 
\begin{align}
T_{e_d,...,e_1}(\{l\},\{a\})=\prod_{f \neq e_d,...,e_1} \frac{1}{l_f\cdot Q+a_f}\ , \label{eq:Quantity}
\end{align}
where $Q$ is the solution of the system
\begin{align}
l_{e_s}\cdot Q+a_{e_s}=0\ , s=1,...,d\ ,
\end{align}
at a generic point in the parameter space. 

For the proof of ~\eqref{eq:tobeproven}, single out a particular coordinate system in $q$-space and apply the identity ~\eqref{eq:Identity} to the component $q_d$, to obtain
\begin{align}
F_N(\{l\},\{a\},q)=\sum_{e_d=1}^N\frac{1}{l_{e_d}\cdot q+a_{e_d}} \prod_{f=1,...,N,\neq e_d}(\frac{1}{l_f\cdot q+a_f} |_{q_d=Q^{e_d}_d(\{l\},\{a\},q_{d-1},...,q_1)})\ ,
\end{align}
where $Q^{e_d}_d(\{l\},\{a\},q_{d-1},...,q_1)$ is the solution of $l_{e_d}\cdot q+a_{e_d}=0$ considered as an equation for $q_d$. The identity ~\eqref{eq:Identity} can be applied to the product under the sum with $q_{d-1}$ singled out, with the result
\begin{align}
F_N(\{l\},\{a\},q)=\sum_{e_d=1}^N\frac{1}{l_{e_d}\cdot q+a_{e_d}} \label{eq:12}\\ \nonumber
& \   & \hspace{-50mm}
\times \sum_{e_{d-1}=1,...,N,\neq e_d}(\frac{1}{l_{e_{d-1}}\cdot q+a_{e_{d-1}}}|_{q_d=Q^{e_d}_d(\{l\},\{a\},q_{d-1},...,q_1)}) \\ \nonumber
& \   & \hspace{-50mm}
\times  \prod_{f=1,...,N,\neq e_d,e_{d-1}}(\frac{1}{l_f\cdot q+a_f} |_{q_s=Q^{e_d,e_{d-1}}_s(\{l\},\{a\},q_{d-2},...,q_1)\ ,s=d,d-1})\ ,
\end{align}
where $Q^{e_d,e_{d-1}}_s\ ,s=d,d-1$ is the solution of the system $l_f\cdot q+a_f=0\ ,f=e_d,e_{d-1}$, considered as equations for $q_d,q_{d-1}$. Continuing in this way, one may arrive at the following identity, valid for any $k, 1\leq k \leq d$,
\begin{align}
F_N(\{l\},\{a\},q)=\sum_{e_d=1}^N\frac{1}{l_{e_d}\cdot q+a_{e_d}}\times \cdot \cdot \cdot \times \label{eq:wow}\\ \nonumber
& \   & \hspace{-50mm}
\times \sum_{e_k=1,...,N,\neq e_d,...,e_{k+1}}(\frac{1}{l_{e_k}\cdot q+a_{e_k}}|_{q_r=Q^{e_d,...,e_{k+1}}_r(\{l\},\{a\},q_k,...,q_1),\ r=d,..,k+1})\\ \nonumber
& \   & \hspace{-60mm}
\times \prod_{f=1,...,N,\neq e_d,...,e_k}(\frac{1}{l_f\cdot q+a_f} |_{q_s=Q^{e_d,...,e_k}_s(\{l\},\{a\},q_{k-1},...,q_1)\ ,s=d,...,k})\ , 
\end{align}
where $Q^{e_d,...,e_k}_r(\{l\},\{a\},q_{k-1},...,q_1),r=d,...,k$ is the solution of the system $l_{e_s}\cdot q+q_{e_s}=0$ for $q_d,...,q_k$.  In particlular, for $k=1$ the product under the iterated sum is independent of $q$. It will be denoted by $T_{e_d,...,e_1}(\{p\},\{a\})$. It coincides with the quantity defined in Eq.~\eqref{eq:Quantity}. It is independent of the order of the indices $e_d,...,e_1$. Thus, the identity ~\eqref{eq:wow} for $k=1$ can be rewritten as
\begin{align}
F_N(\{l\},\{a\},q)=\sum_{f_1\neq... \neq f_d}T_{f_d,...,f_1}(\{p\},\{a\})   \nonumber \\
& \   & \hspace{-70mm}
\times  \sum_{\{e_1,...,e_d\}=\{f_1,...,f_d\}} \frac{1}{l_{e_d}\cdot q+a_{e_d}}\times \cdot \cdot \cdot \times  \nonumber\\ 
& \   & \hspace{-70mm}
\times (\frac{1}{l_{e_1}\cdot q+a_{e_1}}|_{q_r=Q^{e_d,...,e_{2}}_r(\{l\},\{a\},q_1),\ r=d,..,2})  \ , 
\end{align}
where the inner sum extends over permutations of the set $\{f_1,...,f_d\}$. The inner sum is equal to $F_d(\{l_{f_1},...,l_{f_d}\},\{a_{f_1},..,a_{f_d}\},q)$, as can be deduced from the identity ~\eqref{eq:wow} applied to this particular choice of the momenta $l_e$ and scalars $a_e$, with $k=1$, and $N=d$.  This proves the identity ~\eqref{eq:tobeproven}, and hence Eq. \eqref{eq:second}

Combining the two partial fractionings ~\eqref{eq:first},~\eqref{eq:second}, we see that, for $N_q>d$, the product of $N_q\geq d+1$ denominators can be reduced to a sum of terms each of which has only $d+1$ denominators, only one of which is quadratic,
\begin{align}
\prod_e \frac{1}{(q+p_e)^2+m_e^2}  =\sum_e \sum_{j_1,...,j_d} \frac{1}{(q+p_e)^2+m_e^2} \label{eq:parfraconeline} \\   \nonumber
& \   & \hspace{-50mm}
\times \prod_{s=1...d}\frac{1}{2q\cdot(p_{j_s}-p_e)+p_{j_s}^2+m_{j_s}^2-p_e^2-m_e^2} T^e_{j_1,...,j_d}(p_{e'})\ . 
\end{align}

We now embed this result in an arbitrary diagram. The integrand of an arbitrary 1PI Feynman scalar  diagram can be written as
\begin{align}
Integrand= \prod_{i=1}^I \prod_{e_i \in E_i} \frac{1}{(q_i+p_{e_i})^2+m_{e_i}^2}\ .
\end{align}
Here, $i=1 \dots I$ labels the lines of the corresponding bubble diagram and $E_i$ labels the set of lines in the diagram associated with line $i$ in the bubble diagram.   All of the lines in $E_i$ carry the same linear combination of loop momenta, $q_i$, 
\begin{align}
q_i=\sum_{\alpha=1}^{L} l_{i,\alpha}q_\alpha\ ,
\end{align}
where $L$ is the number of loop momenta. $l_{i,\alpha}\in \{ 0,1,-1\}$ can be read from the bubble diagram in the standard way. 

The partial fractioning formula ~\eqref{eq:parfraconeline} can be used to derive the following algebraic identity for the full integrand of any sufficiently dressed scalar integral 
\begin{align}
Integrand  =  \prod_{i\in I} \sum_{e_i \in E_i} \frac{1}{(q_i+p_{e_i})^2+m_{e_i}^2}\sum_{j^i_1,...,j^i_d \in E_i} T^{e_i}_{j^i_1,...,j^i_d}(p_{e'_i}) \label{eq:8}    \nonumber \\%&\ & \hspace{-10mm}  \\ \nonumber
\times \prod_{s=1...d} \frac{1}{2q_i\cdot(p_{j^i_s}-p_{e_i})+(p_{j^i_s})^2+(m_{j^i_s})^2-(p_{e_i})^2-m_{e_i}^2}\ , %\label{eq:integrandpartfrac}
\end{align}
where $T^{e_i}_{j^i_1,...,j^i_d}(p_{e'_i}) $ is obtained through the use of the prescription   of Eq.\  ~\eqref{eq:T} with $p_e,p_j$ substituted accordingly. The order of operations is such that rightmost operations are done first. The argument $p_{e'_i}$ of $T$'s signifies the dependence on other momenta from the line $i$.

The integrals of both sides of Eq.\ ~\eqref{eq:8}  are well-defined on  any representative  of  the  homology class in $H_{dL}(P^{dL}-\cup_i D^c_i )$ ($D^c_i $ stand for the standard ~\cite{hwa} compactifications of the zero sets of the Feynman denominators, and $P^{dL}$ is the complex projective space) that  does not intersect the zero loci of the linear factors. One such choice corresponds to the standard $i\epsilon$ prescription.  Therefore we may extend the equality of integrands in ~\eqref{eq:8} to an equality of integrals, 
\begin{align}
Integral= \int_{\zeta} (dq)  \prod_{i\in I} \sum_{e_i \in E_i} \frac{1}{(q_i+p_{e_i})^2+m_{e_i}^2}\sum_{j^i_1,...,j^i_d} T^{e_i}_{j^i_1,...,j^i_d}(p_{e'_i})  \\ \nonumber
\times \prod_{s=1...d} \frac{1}{2q_i\cdot(p_{j^i_s}-p_{e_i})+(p_{j^i_s})^2+(m_{j^i_s})^2-(p_{e_i})^2-m_{e_i}^2}\ .
\end{align}
It is enough to choose any representative, $\zeta$ of the homology class and deform it in such a way that it does not intersect any of the zero loci of the denominators in the reduction formula. Or one can initially choose such a representative. There are obviously many such choices. On such representatives the reduction formula ~\eqref{eq:8} is valid in each point, so it indeed can be integrated.

Eq.~\eqref{eq:8} for the integral can be rewritten as follows 
\begin{align}
Integral=  \sum_{e_i,j^i_s \in E_i} \prod_{i\in I} T^{e_i}_{j^i_1,...,j^i_d}(p_{e'_i}) \int_{\zeta} (dq) \prod_{i\in I} \frac{1}{(q_i+p_{e_i})^2+m_{e_i}^2} \\   \nonumber
\times \prod_{s=1...d} \frac{1}{2q_i\cdot(p_{j^i_s}-p_{e_i})+(p_{j^i_s})^2+(m_{j^i_s})^2-(p_{e_i})^2-m_{e_i}^2}\ ,
\end{align}
with the single sum over all possible choices of the $d$-tuples of external lines that emanate from each bubble line $i$, in a sufficiently dressed diagram.  The rightmost operations are done first. The indices $j^i_1,...,j^i_d$ of the $T^{e_i}_{j^i_1,...,j^i_d}(p_{e'_i})$ are the same as the indices of the vectors $p_{j^i_s}$  in the second product under the integral sign.

In summary, we see that an arbitrary sufficiently dressed integral can be reduced to a linear combination of integrals of simpler type, with one quadratic factor and exactly $d$ linear factors for each bubble line. It might be that this form of the integral is useful in applications, especially because of the reduced number of poles in each integrand. One must keep in mind that there is still significant freedom in choosing the contour $\zeta$, if one for instance desires to do numeric evaluation of this integral, or relate it to physical amplitudes.

\section{Extension to theories with spin.}

In this section it is shown how to extend the reduction method to theories with arbitrary spinor or vector numerators.  It will be demonstrated that an arbitrary integrand can be reduced to a sum of terms such that there is one quadratic factor per bubble line, up to $d$ linear factors in the denominator and up to $d+1$ linear factors in the numerator. Thus, the conclusion that the number of elements in this basis for integrals grows linearly with the number of loops and is independent of the number of external lines still holds.

We start with the observation that for a product of propagators like the left-hand side of Eq.\ ~\eqref{eq:first}, the number of quadratic factors in the denominator can still be decreased to one through the identity ~\eqref{eq:first} without changing numerator momenta at all. It remains to reduce the products of the form
\begin{align}
I=\frac{ \prod_{e=1}^{E-r} (k_e\cdot q+b_e)}{\prod_{e=1}^{E}(l_e\cdot q+a_e)}\ ,
\end{align}
where $q$ is a loop momentum, $l_e$ are linear combinations of momenta flowing into the line and $k_e$ are arbitrary momenta, not necessarily related to the $l_e$. This generality of the choice for $k_e$ includes open vector indices, so one can choose for instance $k_{e,\mu}=\delta_{\mu\nu}$. The denominator terms $a_e$ are expressed through external momenta and masses,  and the $q$-dependent numerator terms $b_e$ are arbitrary. As above, the integer $E$ is one less the number of propagators that correspond to the bubble line. Integer $r$ is a free parameter of the formula that accounts for the number of linear factors of $q$ in the numerator, ranging from $0$ to $E+1$. The final case can be  realised when all the external lines are attached to the bubble line through a vertex that is linear in momenta, such as a triple gauge boson coupling. Thus, $r=-1,...,E$. 

If $r\geq d$ then a direct generalization of the scalar formula, Eq. ~\eqref{eq:second}, holds
\begin{align}
I= \sum_{f_1,...,f_d} M_{f_1,...,f_d} \frac{1}{\prod_{s=1}^d  (l_{f_s}\cdot q+a_{f_s})}\ , \label{eq:one}
\end{align}
with 
\begin{align}
M_{f_1,...,f_d}=\frac{\prod_{e=1}^{E-r} ( k_e\cdot  Q+b_e)}{  \prod_{e\neq f_1,...,f_d} (l_e\cdot Q+a_e)}\ , \label{eq:M}
\end{align}
where $Q$ is the solution of a system of equations that is analogous to ~\eqref{eq:system} and is constructed from the vectors $k_{f_s}$. Therefore, in this case the basis of intergals is the same as in the scalar case.

In the proof of Eq.\  ~\eqref{eq:one}, the following generalization of the identity ~\eqref{eq:Identity} will be used
\begin{align}
\frac{\prod_{e=1}^F (f_ex+a_e)}{\prod_{e=1}^E (g_ex+b_e)}=\sum_{e=1}^E \frac{1}{g_ex+b_e} (\frac{\prod_{f=1}^F (f_fx+a_f)}{\prod_{f=1,\neq e}^E (g_fx+b_f)})|_{x=-b_e/g_e}\ , \label{eq:newidentity}
\end{align}
which is valid at a generic point in the space of complex variables $f_e,a_e,g_e,b_e,x$, and for all integers $F<E$. The proof involves the same residue argument as before and we omit it. 

The proof of ~\eqref{eq:one} now parallels the proof of ~\eqref{eq:second}. Consider the quantity 
\begin{align}
G_{F,E}(\{l\},\{k\},\{a\},\{b\},q)=\frac{  \prod_{e=1}^F (k_e \cdot q+b_e)}{  \prod_{e=1}^E (l_e \cdot q+a_e)}\ ,
\end{align}
where $l_e,k_e,q \in C^d, a_e,b_e \in C$ and $F\leq E-d$. Application of the identity ~\eqref{eq:newidentity}  to the variable $q_d$ gives
\begin{align}
G_{F,E}(\{l\},\{k\},\{a\},\{b\},q)=\sum_{e_d=1}^E \frac{1}{l_{e_d}\cdot q+a_{e_d}}  \\ \nonumber
& \   & \hspace{-80mm}
 \times (\frac{  \prod_{f=1}^F (k_f \cdot q+b_f)}{  \prod_{f=1,\neq e_d}^E (l_f \cdot q+a_f)}|_{q_d=Q^{e_d}_d})\ .%(\{l\},\{a\},q_{d-1},...,q_1)}\ .
\end{align}
with $Q^{e_d}_d=Q^{e_d}_d(\{l\},\{a\},q_{d-1},...,q_1)$ the solution for $q_d$ of the equation $l_{e_d}\cdot q+a_{e_d}=0$. The identity ~\eqref{eq:newidentity} can be applied again to the product inside the sum, with $q_{d-1}$ taken as $x$. The result, analogous to ~\eqref{eq:12}, is 
\begin{align}
G_{F,E}(\{l\},\{k\},\{a\},\{b\},q)=\sum_{e_d=1}^E \frac{1}{l_{e_d}\cdot q+a_{e_d}} \\ \nonumber
& \   & \hspace{-100mm}
\times \sum_{e_{d-1}=1,\neq e_d}^E (\frac{1}{l_{e_{d-1}}\cdot q+a_{e_{d-1}}}|_{q_d=Q^{e_d}_d})   \\ \nonumber%(\{l\},\{a\},q_{d-1},...,q_1)} \\ \nonumber
& \   & \hspace{-50mm}
 \times\ (\frac{  \prod_{f=1}^F (k_f \cdot q+b_f)}{  \prod_{f=1,\neq e_d,e_{d-1}}^E (l_f \cdot q+a_f)}|_{q_r=Q^{e_d,e_{d-1}}_r, r=d,d-1})\ ,%(\{l\},\{a\},q_{d-2},...,q_1),r=d,d-1}
\end{align}
with $Q^{e_d,e_{d-1}}_r=Q^{e_d,e_{d-1}}_r(\{l\},\{a\},q_{d-2},...,q_1)$ being the solution for $q_d,q_{d-1}$ of the system $l_t\cdot q+a_t=0,t=e_d,e_{d-1}$. After the application of the basic identity ~\eqref{eq:newidentity} $s$ more times the result takes the form
\begin{align}
G_{F,E}(\{l\},\{k\},\{a\},\{b\},q)=\sum_{e_d=1}^E \frac{1}{l_{e_d}\cdot q+a_{e_d}}\times \cdot \cdot \cdot \times \\ \nonumber
& \   & \hspace{-70mm}
\times\ \sum_{e_{d-s}=1,\neq e_d,...,e_{d-s+1}}^E (\frac{1}{l_{e_{d-s+1}}\cdot q+a_{e_{d-s+1}}}|_{q_r=Q^{e_d,...,e_{d-s+1}}_r, ,r=d,...,d-s+1}) \\ \nonumber%(\{l\},\{a\},q_{d-s-1},...,q_1)} \\ \nonumber
& \   & \hspace{-70mm}
 \times\ (\frac{  \prod_{f=1}^F (k_f \cdot q+b_f)}{  \prod_{f=1,\neq e_d,...,e_{d-s}}^E (l_f \cdot q+a_f)}|_{q_r=Q^{e_d,...,e_{d-s}}_r, r=d,...,d-s})\ .%(\{l\},\{a\})}
\end{align}
As in the scalar case for $s=d-1$ can be rewritten as 
\begin{align}
G_{F,E}(\{l\},\{k\},\{a\},\{b\},q)=\sum_{f_1\neq... \neq f_d}M_{f_d,...,f_1}(\{k\},\{l\},\{a\},\{b\})   \nonumber \\
& \   & \hspace{-100mm}
\times  \sum_{\{e_1,...,e_d\}=\{f_1,...,f_d\}} \frac{1}{l_{e_d}\cdot q+a_{e_d}}(\frac{1}{l_{e_{d-1}}\cdot q+a_{e_{d-1}}}|_{q_d=Q^{e_d}_d})\\ \nonumber
\times \cdot \cdot \cdot \times (\frac{1}{l_{e_1}\cdot q+a_{e_1}}|_{q_r=Q^{e_d,...,e_{2}}_r, r=d,..,2})\ ,%(\{l\},\{a\},q_1)}\ , 
\end{align}
with $M$ given by Eq.~\eqref{eq:M}. The sum that multiplies $M_{f_d,...,f_1}$ is precisely $F_d(l_{e_d},...,l_{e_1},a_{e_d},...,a_{e_1},q)$. This proves the identity ~\eqref{eq:one}.

In case $r<d$, $I$ can be represented in the form
\begin{align}
I= \sum_{f_1,...,f_d} M_{f_1,...,f_d} \frac{1}{\prod_{s=1}^d  (l_{f_s}\cdot q+a_{f_s})}\prod_{f=E-d+1,...,E-r} (k_f\cdot q+b_f)\ , \label{eq:30}
\end{align}
with $M$ the same as above, Eq.\ ~\eqref{eq:M}. This proves the assertion made in the beginning of this section. 

Equation ~\eqref{eq:30} can be simplified and recast in the invariant terms, again by partial fractioning. We do it here in the case $r=d-1$, where for Eq.\ ~\eqref{eq:30},
\begin{align}
I= \sum_{f_1,...,f_d} M_{f_1,...,f_d} \frac{(k_{E-d+1}\cdot q+b_{E-d+1})}{\prod_{s=1}^d  (l_{f_s}\cdot q+a_{f_s})}\ .
\end{align}
Momenta $l_{f_s}$ are linearly independent, so $k_{E-d+1}=\sum_s \alpha_s l_{f_s}$ for some $\alpha_s$. Therefore,
\begin{align}
I= \sum_{f_1,...,f_{d-1}} M^{(d-1)}_{f_1,...,f_{d-1}} \frac{1}{\prod_{s=1}^{d-1}  (l_{f_s}\cdot q+a_{f_s})}+\sum_{f_1,...,f_d} M^{(d)}_{f_1,...,f_d} \frac{1}{\prod_{s=1}^d  (l_{f_s}\cdot q+a_{f_s})}\ , \label{eq:two}
\end{align}
for some $M^{(d-1)},M^{(d)}$.

It is possible to determine the coefficients $M^{(d-1)},M^{(d)}$. The $M^{(d)}$ are given by the same formula ~\eqref{eq:M}, as can be seen by studying the asymptotics near $d$-tuple denominator poles. It remains to find $M^{(d-1)}$. In order to do this, consider the line defined by
\begin{align}
l_{f_1}\cdot q+a_{f_1}=0 \label{eq:sys3}\\  \nonumber
....   \\ \nonumber
l_{f_{d-1}}\cdot q+a_{f_{d-1}}=0\ .
\end{align}
Its equation can be written as
\begin{align}
L=\{q_\mu=A_{\mu}+B_{\mu}t | t\in C\}\ .
\end{align}
We multiply Eq. \eqref{eq:two} by the factors $l_{f_1}\cdot q+a_{f_1},...,l_{f_{d-1}}\cdot q+a_{f_{d-1}}$ and consider the result near the line $L$. It is 
\begin{align}
\frac{\prod_{f=1,...,E-d+1} (k_f \cdot (A+Bt)+b_f)}{ \prod_{f \neq f_1,...,f_{d-1}} (l_f\cdot (A+Bt)+a_f)}= M^{(d-1)}_{f_1,..,f_{d-1}} + \sum \frac{M^{(d)}_{f_1,...,f_{d}}}{l_f\cdot (A+Bt)+a_f}\ .
\end{align}
It follows from this formula that
\begin{align}
M^{(d-1)}_{f_1,...,f_{d-1}}= \frac{\prod_{f=1,...,E-d+1} k_f\cdot B}{ \prod_{f \neq f_1,...,f_{d-1}} l_f\cdot B}\ .
\end{align}
Note that $B$ is determined by the $d-1$ hyperplanes of Eq.\ ~\eqref{eq:sys3} up to a constant, which cancels in the above fraction.

\section{Further reduction and symmetries for multiloop integrals. }

In this section further partial fractioning in the linear sector of the integrand is discussed for the example of two-loop scalar integrals. This allows one to reduce the number of linear factors further. Also, a symmetry of the resulting integrals is identified. These techniques can be generalized to higher loops.

In Section 2 it was shown that the scalar integral can be reduced to the following sum
\begin{align}
Integral=\sum_{e_i,j^i_s \in E_i} \prod_i T^{e_i}_{j^i_1,...,j^i_d}(p_{e'_i})  I(p_{e_i},p_{j^i_s},a_{j^i_s})\ ,
\end{align}
where $a_{j^i_s}$ are quadratic functions of masses and momenta that can be read off from the previous formulas. In the two-loop case a generic integral $I$ has the form
\begin{align}
I(p_{e_i},p_{j^i_s},a_{e^i_s})=\int dq_1dq_2 \frac{1}{(q_1+p_{e_1})^2+m_1^2} \frac{1}{(q_2+p_{e_2})^2+m_2^2} \\ \nonumber
\times  \frac{1}{(q_1+q_2+p_{e_3})^2+m_3^2} \prod_{j^1_s \in E_1} \frac{1}{q_1\cdot (p_{j^1_s}-p_{e_1})+a_{j^1_s}} \\ \nonumber
\times \prod_{j^2_s \in E_2}  \frac{1}{q_2\cdot (p_{j^2_s}-p_{e_2})+a_{j^2_s}} \prod_{j^3_s \in E_3}  \frac{1}{(q_1+q_2)\cdot (p_{j^3_s}-p_{e_3})+a_{j^3_s}}\ ,
\end{align}
where $i \in 1,2,3$, $E_i$ are $d$-element sets of the indices $j^i_s$, and $a_{j^i_s}$ are functions of masses and momenta that one can read off from Eq.\ (10). The change of variables 
\begin{align}
q'_1  =  q_1+p_{e_1}\ , \\ \nonumber
q'_2=  q_2+p_{e_2}\ ,
\end{align}
will eliminate the dependence of the two quadratic factors on the external momenta. The prime in $q'_i$ will be suppressed in the following. In order to simplify the notation, the following change of coordiantes can be made 
\begin{eqnarray}
r & = & p_{e_3}-p_{e_1}-p_{e_2}\ , \\ \nonumber
p_{s_i} & = & p_{j^i_s}-p_{e_i}\ , a_{s_i}=a_{j^i_s}-p_{s_i}\cdot p_{e_i}\ ,\quad  s_i \in E_i\ ,i=1,2,3\ .
\end{eqnarray}
In these variables, the integral $I$, which will still be denoted by the same letter, assumes the form
\begin{align}
I(p_{e_i},p_{j^i_s},a_{j^i_s})=\int dq_1dq_2\ \frac{1}{q_1^2+m_1^2} \frac{1}{q_2^2+m_2^2} \frac{1}{(q_1+q_2+r)^2+m_3^2} \label{eq:41}\\ \nonumber
\times \prod_{s_1 \in E_1} \frac{1}{q_1\cdot p_{s_1}+a_{s_1}} \prod_{s_2 \in E_2}  \frac{1}{q_2\cdot p_{s_2}+a_{s_2}} \prod_{s_3 \in E_3}  \frac{1}{(q_1+q_2+r)\cdot p_{s_3}+a_{s_3}}\ .
\end{align}

%\subsection{Fractioning in $q_2$-sector.}

In the general two-loop integral, Eq.\ ~\eqref{eq:41}, there  are exactly $2d$ linear factors that have $q_2$ dependence, namely
\begin{align}
P_2=  \prod_{s_2\in E_2} \frac{1}{q_2\cdot p_{s_2}+a_{s_2}} \prod_{s_3 \in E_3}  \frac{1}{(q_1+q_2+r)\cdot p_{s_3}+a_{s_3}}\ . \label{eq:origp2}
\end{align}
where $E_i$ are $d$-element sets of indices. The same partial fractioning as in Eq.\ (2) for products of linear denominators can be done on these factors with the result 
\begin{align}
P_2=\sum_{|E'_2|+|E'_3|=d} T'_{E'_2,E'_3} \prod_{s_2\in E'_2} \frac{1}{q_2\cdot p_{s_2}+a_{s_2}} \prod_{s_3 \in E'_3} \frac{1}{(q_1+q_2+r)\cdot p_{s_3}+a_{s_3}}\ , \label{eq:p2}
\end{align}
where we have introduced the subsets $E_1'$ and $E_2'$ whose union has $d$ elements, as indicated by the first summation over all such subsets.   The functions $T'_{E_1',E_2'}$ are the analogs of the functions $T$ in Eq.\ (2),
\begin{align}
T'_{E'_2,E'_3}=\prod_{s_2 \notin E'_2} \frac{1}{Q'\cdot p_{s_2}+a_{s_2}} \prod_{s_3 \notin E'_3} \frac{1}{(Q'+q_1+r)\cdot p_{s_3}+a_{s_3}}\ , \label{eq:dindon}
\end{align}
where $Q'$ is the solution of the following linear system
\begin{eqnarray}
Q'\cdot p_{s_2}+a_{s_2} & = & 0, s_2 \in E'_2  \label{eq:bombom}\label{eq:45}\\ \nonumber
(Q'+q_1+r)\cdot p_{s_3}+a_{s_3} & = & 0, s_3 \in E'_3\ .
\end{eqnarray}
The solution of this linear sysytem can be found by inverting the $d\times d$ matrix whose rows are the $d$ vectors $p_{s_2}$ and $p_{s_3}$ in Eq.\ ~\eqref{eq:bombom}, which we assume to be linearly independent.  It can thus be written 
\begin{align}
Q'_\mu=f_\mu+e_{\mu \nu}q_{1,\nu}\ , \label{eq:solution}
\end{align}
where $f,e$ are rational functions of all external momenta and are independent of $q_1$. The functions $T'_{E'_2,E'_3}$ in Eq.\ ~\eqref{eq:dindon} can thus be  written as
\begin{align}
T'_{E_2',E_3'} = \prod_{f_i\in F} \frac{1}{q_1\cdot  v_{f_i} + d_{f_i}}\ ,
\end{align}
where $F=(E_2-E'_2) \cup (E_3 - E'_3)$ is a $d$-element set. We thus have from Eq. ~\eqref{eq:p2},
\begin{align}
P_2=\sum_{|E'_2|+|E'_3|=d} \prod_{f_i\in F} \frac{1}{q_1\cdot  v_{f_i} + d_{f_i}} \prod_{s_2\in E'_2} \frac{1}{q_2\cdot p_{s_2}+a_{s_2}} \prod_{s_3 \in E'_3} \frac{1}{(q_1+q_2+r)\cdot p_{s_3}+a_{s_3}}\ . \label{eq:newp2}
\end{align}
Explicit expressions for $v$ and $d$ are obtained by substituting the solutions ~\eqref{eq:solution} in Eq.\ ~\eqref{eq:dindon}.  In the resulting form, the 2$d$ $q_2$-dependent denominators $P_2$ in Eq. ~\eqref{eq:p2} are transformed to a sum of terms with $d$ linear factors in $q_1$ and  $q_2$, times $d$ linear factors in $q_1$ alone.  So far, the total number of $q_i$-dependent denominators is still the same as in Eq.\ ~\eqref{eq:p2}.

Partial fractioning can now be applied to the $2d$ remaining denominators that depend on $q_1$ but not $q_2$,            
\begin{eqnarray}
P_1&=&\prod_{s_1 \in E_1} \frac{1}{q_1\cdot p_{s_1}+a_{s_1}} T'_{E'_2,E'_3} \label{eq:opa}\\ \nonumber
&=&\prod_{s_1 \in E_1} \frac{1}{q_1\cdot p_{s_1}+a_{s_1}} \prod_{f_i\in F} \frac{1}{q_1\cdot  v_{f_i} + d_{f_i}}\ ,
\end{eqnarray}
where $F$ is the $d$-element set in ~\eqref{eq:newp2}. Going through the same steps of  partial fractioning that led to ~\eqref{eq:p2} this quantity can be reduced to a linear combination of $d$ linear factors in $q_1$ times products of fractions that depend only on external momenta,
\begin{align}
P_1=\sum_{E'_1} T_{E'_1}(p,a)\prod_{s_1 \in E'_1}\frac{1}{q_1 \cdot \tilde p_{s_1}+\tilde a_{s_1}}\ ,
\end{align}
where $T_{E'_1}(p,a)$ are the familiar $T$-functions, which in this case are obtained from partial fractioning of ~\eqref{eq:opa}, and the solution of a system of equations, as in ~\eqref{eq:45}.

Thus, the original integral $I$, Eq. \eqref{eq:origp2}, can be reduced to linear combinations of a new set of integrals with $2d$ loop momentum-dependent denominators, reduced from $3d$,
\begin{align}
I(p_{e_i},p_{j^i_s},a_{j^i_s}) = \sum_{ |E'_1| = d,|E'_2|+|E'_3|=d} I'(r,\tilde p_{s_i},\tilde a_{s_i}) \ T_{E'_1}(p,a)\ ,
\end{align}
where $T_{E'_1}(p,a)$ is a function of external momenta only and where
\begin{align}
I'(r,\tilde p_{s_i},\tilde a_{s_i})=\int dq_1dq_2 \frac{1}{q_1^2+m_1^2} \frac{1}{q_2^2+m_2^2} \frac{1}{(q_1+q_2+r)^2+m_3^2} \label{eq:24} \\ \nonumber
\times \prod_{s_1 \in E'_1} \frac{1}{q_1\cdot \tilde p_{s_1}+\tilde a_{s_1}} \prod_{s_2 \in E'_2}  \frac{1}{q_2\cdot p_{s_2}+a_{s_2}} \prod_{s_3 \in E'_3}  \frac{1}{(q_1+q_2+r)\cdot p_{s_3}+a_{s_3}}\ .
\end{align}
This looks exactly like the original integrals ~\eqref{eq:41} with the exception that now the sets $E'_2,E'_3$ in total have $d$ elements (while in the original integral they had $2d$ elements). Quantities $m,r,p_{s_2},a_{s_2},p_{s_3},a_{s_3}$ are the same as in the original integral, while the vectors $\tilde p_{s_1}$ and the scalars $\tilde a_{s_1}$ are rational functions of the variables of the original integral found by solving linear systems of equations, like ~\eqref{eq:bombom}. Note that the intergal $I'$ is still overall UV-convergent, and UV-convergent in both $q_1,q_2$ sectors.

%I'(p_{e_i},p'_{j^i_s},a'_{j^i_s})=\int dq_1dq_2 \frac{1}{q_1^2+m_1^2} \frac{1}{q_2^2+m_2^2} \frac{1}{(q_1+q_2+p_{12})^2+m_3^2} \\ \nonumber
%\times \prod_{e_i \in E'_i} \frac{1}{q_1\cdot p'_{e_1}+a'_{e_1}}  \frac{1}{q_2\cdot p_{e_2}+a_{e_2}}  \frac{1}{(q_1+q_2+s)\cdot p_{e_3}+a_{e_3}}\ ,

%\subsection{Symmetry under the change of variables.}

Finally, under the change of variables
\begin{align}
q_2'=q_2+q_1+r\\ \nonumber
q_1'=-q_1
\end{align}
the function $I'$ transforms as
\begin{align}
I'(r;p_{s_1},a_{s_1};p_{s_2},a_{s_2}; p_{s_3},a_{s_3}) = I'(-r;-p_{s_1},a_{s_1}; p_{s_3},a_{s_3};p_{s_2},a_{s_2})\ ,
\end{align}
where $s_1 \in E'_1, s_2 \in E'_2, s_3 \in E'_3$. This identity allows the integral to be reduced to a form with no more than $[d/2]$ linear factors with $q_1+q_2$ , where $[x]$ is the largest integer not greater than $x$. Indeed, the integrals $I'$ have the property that $|E'_2|+|E'_3|=d$. Therefore, if originally $|E'_2| \geq [d/2]$, then after such change of varibales $|E'_2| \leq [d/2]$.

\section{Conclusion.}

This paper describes a new method for the simplification of one-particle irreducible perturbative diagrams with many external lines.  Each set of internal lines that carry the same linear combination of loop momentum in any diagram is rewritten as a sum of terms with one quadratic and no more that $d$ linear denominators in that momentum.  In the case of tensor integrals, there can be up to $d+1$ linear factors in the numerator as well.  This reduction is achieved through the use of two elementary identities, both of which are variants of partial fractioning, and which are analogous to propagator identities described in Refs.\ \cite{Vaman:2005dt},\cite{Draggiotis:2005wq} for tree diagrams.    Arbitrary integrals are linear combinations of a basis integrals of this kind, with coefficients that are rational functions of external momenta and masses, which may take on arbitrary complex values.  The analytic continuation of correlation functions is thus simplified.

The results of this study may offer a new starting point for studies of the analytic properties of correlation functions in perturbation theory, related to the use of the unitarity method \cite{Britto:2010xq},\cite{Dixon:2011xs}.  By writing complex diagrams as sums of integrals with reduced numbers of poles, it may be possible to simplify the study of both asymptotic behavior and analytic structure.  These applications will be the subject of future research.

%\section{Appendix:proof of the partial fractioning formula for linear factors.}

\subsection*{Acknowldgements}

I thank George Sterman for discussions and encouragement.
This work was supported in part by the National Science Foundation, grant PHY-0969739.

\end{document}